\documentclass[a4paper]{article}
\usepackage{amssymb}        
\usepackage{amsmath}        
\usepackage{graphicx}
\usepackage{epsfig}        
\usepackage{psfrag}

\title{On the initial-value problem of the Maxwell-Lorentz equations}
\author{Volker Perlick\thanks{Physics Department, Lancaster University, 
Lancaster LA1 4YB, United Kingdom, and The Cockcroft Institute,
Warrington WA4 4AD, United Kingdom. Email: v.perlick@lancaster.ac.uk}
{ and } 
Anthony Carr\thanks{Physics Department, Lancaster University, 
Lancaster LA1 4YB, United Kingdom. Email: carranty@gmail.com}
}

\date{}

\begin{document}

\maketitle

%---------------------------------------------------------------------
\begin{abstract}
\noindent
We consider the Maxwell-Lorentz equations, i.e., the equation of motion 
of a charged dust coupled to Maxwell's equations, on an arbitrary 
general-relativistic spacetime. We decompose this system of equations 
into evolution equations and constraints, and we demonstrate that the 
evolution equations are strongly hyperbolic. This result guarantees that 
the initial-value problem of the Maxwell-Lorentz equations is well-posed.
We illustrate this general result with a discussion of spherically symmetric
solutions on Minkowski spacetime.
%\\[0.2cm]
%PACS: 41.20 -q, 52.30 Ex.
\end{abstract}

%---------------------------------------------------------------------------------------------------
\section{Introduction}\label{sec:intro}

In this article we consider the equations of motion for a charged dust
coupled to an electromagnetic field on a general-relativistic spacetime.
These equations of motion are often refered to as the Maxwell-Lorentz 
equations, because they comprise the Maxwell equations for the 
electromagnetic field and the Lorentz force equation for the dust 
particles. Here, as usual, we use the word ``dust'' as synonymous to 
``pressure-less fluid''. As an alternative, the name ``cold fluid''
is sometimes used. 

The Maxwell-Lorentz equations have several interesting applications. E.g., 
they can be used for describing the electron component in a plasma. Among 
other things, this is of relevance for astrophysics, e.g., for modeling the 
Solar corona. In spite of the high temperature of the Solar corona, it is 
quite reasonable to model the electron component as a pressure-less (or 
``cold'') fluid, because the density is low. The Maxwell-Lorentz equations have 
also been suggested for modeling a beam of charged particles in an accelerator, 
cf. Burton, Gratus and Tucker \cite{BurtonGratusTucker2007}. Special interest 
has been given to the Maxwell-Lorentz equations for \emph{rigid} charge 
distributions on Minkowski spacetime. The idea is to use this as a model 
for an extended charged classical particle, with the self-interaction 
taken into account. Interesting results have been found, e.g., by Spohn 
and collaborators \cite{Spohn2004,KomechSpohn2000,KunzeSpohn2000} as
well as by Bauer and D{\"u}rr \cite{BauerDuerr2001}. An important goal 
of this line of research is to get some new insight into the point-particle
limit and its notorious pathologies. 

Several fundamental features of the Maxwell-Lorentz equations on Minkowski 
spacetime are discussed in Chapter 5 of Parrott's text-book on electrodynamics 
\cite{Parrott1987}. Among other things, Parrott discusses the initial-value 
problem for these equations. Using the Cauchy-Kowalevski theorem, he is 
able to demonstrate local existence and uniqueness of (analytic) solutions 
from analytic initial data. However, the assumption of analyticity 
is physically not quite satisfactory: Knowing an analytic function on an 
arbitrarily small neighborhood fixes the function on the whole domain of 
convergence of the Taylor series, whereas physical measurements in an
arbitrarily small neighborhood cannot determine any results outside 
of the causal future of this neighborhood. Also, it is often convenient
to consider fields which are zero on an open subset $\mathcal{U}$ and
non-zero on the complement of $\overline{\mathcal{U}}$; such fields cannot
be analytic near the boundary of $\mathcal{U}$. For these reasons it is 
desirable to have an existence and uniqueness theorem for non-analytic 
data, as noted by Parrott himself. It is the main purpose of this paper 
to establish such a result. 

To that end we decompose the Maxwell-Lorentz equations, on an arbitrary
general-relativistic spacetime, into evolution equations and constraints.
The evolution equations can be written as a quasi-linear system of first
order partial differential equations. We derive the characteristic equation
(dispersion relation), and we demonstrate that the system of evolution
equations is strongly hyperbolic, which is a necessary and sufficient 
condition for the initial-value problem to be well-posed. This means
that a solution locally exists, that it is unique, and that it depends
continuously on the data in an appropriate topology. This result is of 
relevance for solving the Maxwell-Lorentz equations numerically: If a
solution does not exist it is pointless to write a numerical code that 
tries to approximate it; if the solution is not unique the code does
not know which solution to approximate; and if the dependence on the
data is discontinuous it is impossible to estimate the error of a 
numerical solution. 

After establishing our general existence and uniqueness result, we 
illustrate it with a discussion of spherically symmetric solutions on 
Minkowski spacetime.

%---------------------------------------------------------------------------------------------------
\section{The Maxwell-Lorentz equations}\label{sec:maxlor}

We consider the coupled equations of motion of a charged dust and
an electromagnetic field on a general-relativistic spacetime, i.e., on
a 4-dimensional manifold $M$ with a Lorentzian metric $g$. We
allow for an external electric current density $J^{e}$, and we
assume that the dust consists of only one species of particles, with
charge $q$ and mass $m$. In the following we assume that the second-rank 
tensor field $g=g_{ij} dx^i \otimes dx^j$, the one-form $J^e= J_i^e
dx^i$ and the constants $q$ and $m$ are given. We use standard index
notation with Einstein's summation convention for latin indices taking 
values $0,1,2,3$ and for greek indices taking values $1,2,3$. As usual,
the contravariant metric components are defined by $g^{ki}g{}_{ij} = 
\delta ^k _j$. The metric has signature $(-,+,+,+)$ and its Levi-Civita 
connection is denoted $\nabla$. We use units making $\varepsilon _0$ 
and $\mu _0$ (and, thus, the vacuum speed of light $c$) equal to 1.

The equations of motion consist of Maxwell's equations for the electromagnetic
field and the Lorentz force equation for the dust particles,
\begin{gather}\label{eq:Max1}
\partial _i F_{jk} \, + \,  
\partial _{j} F_{ki} \, + \, 
\partial _{k} F_{ij}
\,= \, 0 \; ,
\\[0.2cm]
\label{eq:Max2}
g^{ij} \nabla _{i} F_{jk} \, = \, \, J_k ^e \, - \, q \, n \, g_{k \ell}
U^{\ell} \; ,
\\[0.2cm]
\label{eq:Lor}
U^i \nabla _i U^j \, = \, \dfrac{q}{m} \, g^{jk}
F_{k \ell} U^{\ell} \; .
\end{gather}
We refer to these equations as to the \emph{Maxwell-Lorentz equations}. The
field variables (i.e., the ``unknowns'' in these equations) are
\begin{itemize}
\item
the electromagnetic field strength $F_{ij} dx^i \otimes dx^j$
which is an antisymmetric second-rank tensor field, $F_{ij} = -F_{ji} \,$;
\item
the four-velocity $U^i \partial _i$ of the dust which is a vector 
field, normalised to $g_{ij}U^i U^j = - 1 \,$.
\item
the (proper) number density $n$ of the dust which is a non-negative scalar function;
\end{itemize}
The electromagnetic field is to be interpreted as the field produced by the
external current plus the field produced by the charged dust plus, possibly,
an additional source-free field.

If $n=0$ on an open subset of the spacetime, $U^i$ is physically
undetermined on this subset. In this case, the electromagnetic field is 
determined by  (\ref{eq:Max1}) and (\ref{eq:Max2}), i.e., by
Maxwell's equations with a given source $J_k^e$. As the initial-value
problem for this system of equations is well-known, we assume in
the following that $n>0$ on our spacetime manifold. (Or, to put this
another way, we restrict our spacetime to an open subset on
which $n>0$.)

Note that (\ref{eq:Lor}) implies $U^i \partial _i \big( g_{jk}U^j U^k 
\big) = 0$. This guarantees that, if the condition $g_{jk}U^jU^k=-1$ 
is satisfied at one point, then it is satisfied along the whole integral 
curve of $U^i \partial _i$ that passes through this point. A similar 
statement is true for the condition $n > 0$, provided that the external 
current is conserved, $g^{ik} \nabla _i J_k^e =0$. To prove this, we 
read from (\ref{eq:Max2}) that the conservation law for the external 
current implies $\nabla _k  \big( nU^k \big) =0$ and thus
\begin{equation}\label{eq:nconti}
U^k \partial _k n + n \nabla _k U^k = 0 \; .
\end{equation}
If $x(t)$ is an integral curve of the vector field $U^i \partial _i$, 
integration of (\ref{eq:nconti}) yields 
\begin{equation}\label{eq:nint}
n \big( x(t_2) \big) \, = \, n \big( x(t_1) \big) \,\mathrm{exp} \Big\{
- \int _{t_1} ^{t_2} \big( \nabla _k U^k \big) \big( x(t) \big) \, dt \Big\}
\end{equation}
so $n$ cannot change sign along an integral curve of $U^i \partial _i$. 
This is an important observation in view of the initial value problem. It
makes sure that, for a solution of the Maxwell-Lorentz equations, it 
suffices to check the conditions $g_{ij}U^iU^j=-1$ and $n > 0$ on 
the initial hypersurface. 

To illustrate the physical content of the Maxwell-Lorentz equations we
briefly mention two different applications. Firstly, we can use these
equations for modeling plasma waves in the Solar corona. In this case we
would choose for $(M,g)$ the Schwarzschild spacetime, for $J^e$ the electric
current density of the ion component which could be assumed static, and 
for $q$ and $m$ the charge and the mass of the electron. The field 
variables are the total electromagnetic field,  the four-velocity of the 
electrons and the number density of the electrons. Secondly, we can 
consider an electron beam in an accelerator. Then we would choose for 
$(M,g)$ an open subset of the Minkowski spacetime (to model the 
spacetime region inside the accelerator pipe), for $J^e$ the zero 
one-form (as there is no external current inside the accelerator pipe), 
and for $q$ and $m$ the charge and the mass of the electron. The 
field variables are again the total electromagnetic field (i.e., the field 
produced by the dust and the accelerating field which has no sources 
inside the pipe), the four-velocity of the electrons and the number density 
of the electrons. 

%---------------------------------------------------------------------------------------------------
\section{Decomposition of the Maxwell-Lorentz equations into evolution 
equations and constraints}\label{sec:evolution}

We fix a spacetime point $P$ and we choose local coordinates, on
an appropriate neighborhood of $P$, such that the
hypersurfaces $x^0= \,$constant are spacelike and the $x^0-$lines are
perpendicular to these hypersurfaces. We want to discuss if the
prescription of appropriate initial data on the hypersurface 
$x^0= \,$constant through $P$ determines a unique solution of the
Maxwell-Lorentz equations on some neighborhood of $P$. To that 
end we have to decompose the Maxwell-Lorentz equations into 
evolution equations (i.e., equations that do contain $\partial _0$
derivatives) and constraints (i.e., equations that do not contain
$\partial _0$ derivatives).

There are two constraints, given by equation (\ref{eq:Max1}) with
all three indices spatial and by equation (\ref{eq:Max2}) with $k = 0$.
After expressing covariant derivatives in terms of partial derivatives,
introducing the Christoffel symbols of the metric tensor, these two
equations take the following form.
\begin{gather}\label{eq:con1}
\partial _{\mu} F_{\nu \sigma} \, + \,  
\partial _{\nu} F_{\sigma \mu} \, + \, 
\partial _{\sigma} F_{\mu \nu}
\,= \, 0 \; ,
\\[0.2cm]
\label{eq:con2}
g^{\mu \nu } \big( \partial _{\mu} F_{\nu 0} - 
\Gamma _{\mu \nu}^{\sigma} F_{\sigma 0}-\Gamma _{\mu 0}^{\sigma}F_{\nu \sigma}
-\Gamma _{\mu 0}^0F_{\nu 0} \big) \, = \, 
J_0^e  \, - \, q \, n \, g_{00}U^0 \; .
\end{gather}
The remaining equations are the evolution equations: Evaluating equation (\ref{eq:Max1}) 
with one of the indices equal to 0 gives three scalar equations,
\begin{equation}\label{eq:evo1}
\partial _0 F_{\mu \nu} \, + \,  
\partial _{\mu} F_{\nu 0} \, + \, 
\partial _{\nu} F_{0 \mu}
\, = \, 0 \; .
\end{equation} 
Evaluating equation (\ref{eq:Max2}) with a spatial index, 
$k = \sigma$, gives another three scalar equations. After 
eliminating the number density $n$, with the help of (\ref{eq:con2}),
these equations read
\begin{gather}
\nonumber
\partial _0 F_{0 \sigma} \, + \, \dfrac{g^{\mu \nu}}{g^{00}} \, \partial _{\mu} F_{\nu \sigma}
\, - \, \dfrac{U^{\rho}}{U^0} \, g_{\sigma \rho}g^{\mu \nu} \partial _{\mu} F_{\nu 0} 
\, = \, \dfrac{g^{\mu \nu}}{g^{00}} \, \big( \Gamma _{\mu \nu}^0 F_{0 \sigma} + 
\Gamma _{\mu \nu} ^{\rho} F_{\rho \sigma} + \Gamma _{\mu \sigma}^0 F_{\nu 0} + 
\Gamma _{\mu \sigma}^{\rho} F_{\nu \rho} \big)
\\[0.2cm]
\label{eq:evo2}
\, - \, \dfrac{U^{\rho}}{U^0} \, g_{\sigma \rho} g^{\mu \nu} 
\big( \Gamma _{\mu \nu}^{\tau} F_{\tau 0} +
\Gamma _{\mu 0}^{\rho} F_{\nu \rho} + \Gamma _{\mu 0}^0 F_{\nu 0} \big) 
\, + \, \Gamma _{00}^0 F_{0 \sigma}
\, + \, \Gamma _{00}^{\mu}F_{\mu \sigma} \, + \, \Gamma _{0 \sigma}^{\mu}F_{0 \mu}
\, + \, \dfrac{J_{\sigma} ^e}{g^{00}} \, - \, 
\dfrac{U^{\rho}}{U^0} \, g_{\sigma \rho} \, J_0^e  \; .
\\[0.15cm]
\nonumber
\end{gather}
Finally, evaluating equation (\ref{eq:Lor}) with a spatial index,
$j = \sigma$, gives us again three scalar equations. With the help of equation 
(\ref{eq:Lor}) with $j=0$, these equations can be rewritten as
\begin{gather}
\nonumber
\partial _0  \dfrac{U^{\sigma}}{U^0}  \, + \, \dfrac{U^{\rho}}{U^0} \partial _{\rho}
\dfrac{U^{\sigma}}{U^0}
\, = \, - \,
\Gamma _{00}^{\sigma} \, - \, 2 \Gamma _{0 \mu}^{\sigma} \, \dfrac{U^{\mu}}{U^0} \, - \, 
\Gamma _{\mu \nu}^{\sigma} \, \dfrac{U^{\mu} U^{\nu}}{(U^0)^2}  
\\[0.15cm]
\label{eq:evo3}
+ \, \dfrac{q}{m} \, \dfrac{1}{U^0} \, \Big\{ g^{\sigma \tau}
\Big( \, F_{\tau 0} \, + \, F_{\tau \rho} \, \dfrac{U^{\rho}}{U^0} \, \Big) \, - \, 
\dfrac{U^{\sigma}U^{\tau}}{(U^0)^2} \, g^{00} \, F_{0 \tau} \, \Big\} \, 
\; .
\end{gather}
Equation (\ref{eq:Lor}) with $j = 0$ can be ignored because it is 
automatically satisfied if the other equations are, owing to the 
normalisation condition $g_{ij}U^i U^j=-1 \,$. So the system of
evolution equations is given by nine scalar equations, (\ref{eq:evo1}),
(\ref{eq:evo2}) and (\ref{eq:evo3}). Note that the number density 
$n$ is not a dynamical variable; $n$ has been eliminated
from the evolution equations and is algebraically determined 
by the other field variables via (\ref{eq:con2}).

To write the evolution equations in a more convenient form, we decompose the 
electromagnetic field into its electric and magnetic components, $E_{\sigma}$ 
and $B^{\nu}$, with respect to the chosen coordinate system, 
\begin{gather}\label{eq:EB}
F_{\sigma 0} \, = \, E_{\sigma} \qquad \mathrm{and} \qquad 
F_{\tau \rho} \, = \, \varepsilon _{\tau \rho \nu} B^{\nu} \; .
\end{gather}
Here $\varepsilon _{\sigma \rho \nu}$ is the three-dimensional covariant 
$\varepsilon$--symbol, defined by being totally antisymmetric and satisfying 
$\varepsilon _{123} =1$. Moreover, we introduce the 3-velocity $v^{\sigma}$ 
of the dust,
\begin{gather}\label{eq:v}
U^i \partial _i \, = \, U^0 \, 
\big( \partial _0  +  v^{\sigma} \partial _{\sigma} \big) \; ,
\end{gather}
Note that then $U^0$ is determined by the normalisation condition $(U^0)^2 (g_{00}
+ v^{\mu}v^{\nu}g_{\mu \nu}) = - 1 \,$. If we use the three-dimensional contravariant 
$\varepsilon$--symbol $\varepsilon ^{\sigma \rho \nu}$, defined by being totally 
antisymmetric and satisfying $\varepsilon ^{123}=1$, we can write the evolution 
equations in the following final form. (We use the identity  $\varepsilon 
^{\mu \nu \sigma} \varepsilon _{\mu \nu \tau} = 2 \delta ^{\sigma} _{\tau}$.)
\begin{gather}\label{eq:evo4}
\partial _0 B^{\sigma} + \varepsilon ^{\sigma \mu \nu} \partial _{\mu} E_{\nu} 
\, = \, 0 \; ,
\\[0.6cm]
\nonumber
\partial _0 E_{\sigma} \, - \, \dfrac{g^{\mu \nu}}{g^{00}} \, \varepsilon _{\nu \sigma \tau}
\partial _{\mu} B^{\tau} \, + \, 
v^{\rho} \, g_{\sigma \rho} g^{\mu \nu} \partial _{\mu} E_{\nu} 
\, = \, 
\dfrac{g^{\mu \nu}}{g^{00}} \, \big( \Gamma _{\mu \nu}^0 E_{\sigma} - 
\Gamma _{\mu \nu} ^{\rho} \varepsilon _{\rho \sigma \nu}B^{\nu} - 
\Gamma _{\mu \sigma}^0 E_{\nu} - 
\Gamma _{\mu \sigma}^{\rho} \varepsilon _{\nu \rho \tau} B^{\tau} \big)
\\[0.15cm]
\label{eq:evo5}
+ \, v^{\rho} \, g_{\sigma \rho} g^{\mu \nu} \big( \Gamma _{\mu \nu}^{\tau} E_{\tau} +
\Gamma _{\mu 0}^{\rho} \varepsilon _{\nu \rho \lambda} B^{\lambda} + 
\Gamma _{\mu 0}^0 E_{\nu} \big) \, - \, \dfrac{J_{\sigma}^e}{g^{00}} \, + \, 
v^{\rho} \, g_{\sigma \rho} \, J_0^e \, + \, \Gamma _{00}^0 E_{\sigma}
\, - \, \Gamma _{00}^{\mu} \varepsilon _{\mu \sigma \nu} B^{\nu} \, + \, 
\Gamma _{0 \sigma}^{\mu} E_{\mu} \; ,
\\[0.6cm]
\nonumber
\partial _0  v^{\sigma}  \, + \, v^{\rho} \partial _{\rho}
v^{\sigma} \, = \, - \,
\Gamma _{00}^{\sigma} \, - \, 2 \Gamma _{0 \mu}^{\sigma} \, v^{\mu} \, - \, 
\Gamma _{\mu \nu}^{\sigma} \, v^{\mu} v^{\nu}  
\\[0.15cm]
\label{eq:evo6}
 + \, \dfrac{q}{m} \, 
\sqrt{-g_{00}-v^{\mu}v^{\nu}g_{\mu \nu}} \,
\Big\{ g^{\sigma \tau}
\Big( \, E_{\tau} \, + \, 
\varepsilon _{\tau \rho \lambda} v^{\rho} B^{\lambda} \, \Big) 
\, + \, g^{00}  v^{\sigma} v^{\tau}  E_{\tau} \, \Big\}
\; .
\end{gather}
With the same notation, the constraints (\ref{eq:con1}) and 
(\ref{eq:con2}) read
\begin{gather}\label{eq:con3}
\partial _{\mu} B^{\mu} \, = \, 0 \; ,
\\[0.2cm]
\label{eq:con4}
q \, n \, = \, \dfrac{\sqrt{-g_{00}-v^{\lambda}v^{\rho}g_{\lambda \rho}}}{- \, g_{00}}
\, \Big( g^{\mu \nu} \big(\partial _{\mu} E_{\nu} - 
\Gamma _{\mu \nu}^{\sigma} E_{\sigma} -
\Gamma _{\mu 0}^{\sigma} \, \varepsilon _{\nu \sigma \tau} B^{\tau} -
\Gamma _{\nu 0}^0 E_{\mu} \big) - J_0^e \, \Big) \; .
\end{gather}
Note that (\ref{eq:evo4}) implies $\partial _0  \partial _{\sigma}
B^{\sigma} = 0$. Hence a solution of the evolution equations satisfies 
(\ref{eq:con3}) everywhere if it satisfies (\ref{eq:con3}) on the initial 
hypersurface $x^0 = \;$constant. As a consequence, the solutions to the
Maxwell-Lorentz equations can be found by solving the initial-value
problem of the evolution equations, with initial values for $E_{\sigma}$,
$B^{\mu}$ and $v^{\rho}$ that satisfy (\ref{eq:con3}) and that give 
a non-negative number density $n$ via (\ref{eq:con4}). Recall that,
by (\ref{eq:nint}), it suffices to check the condition of $n$ being 
positive on the initial hypersurface.

If we write 
\begin{equation}\label{eq:BEv}
\vec{E} \, = \, 
\begin{pmatrix}
E_1 \\ E_2 \\ E_3
\end{pmatrix}
\: , \qquad
\vec{B} \, = \, 
\begin{pmatrix}
B^1 \\ B^2 \\ B^3
\end{pmatrix}
\: , \qquad
\vec{v} \, = \, 
\begin{pmatrix}
v^1 \\ v^2 \\ v^3
\end{pmatrix}
\: ,
\end{equation}
the evolution equations (\ref{eq:evo4}), (\ref{eq:evo5}) and 
(\ref{eq:evo6}) can be written in matrix form as a differential
equation for a 9-component column vector,
\begin{equation}\label{eq:evovec}
\partial _0 \,
\begin{pmatrix}
\vec{E} \\ \vec{B} \\ \vec{v}
\end{pmatrix}
\, - \, 
\boldsymbol{L}^{\rho} \partial _{\rho}
\begin{pmatrix}
\vec{E} \\ \vec{B} \\ \vec{v}
\end{pmatrix}
\, = \, 
\begin{pmatrix}
\vec{m}{}_1 \\ \vec{m}{}_2 \\ \vec{m}{}_3
\end{pmatrix}
\; .
\end{equation}
Here, $\boldsymbol{L}^1 , \boldsymbol{L}^2 , \boldsymbol{L}^3$ are
$9\times 9$ matrices and $\vec{m}{}_1$, $\vec{m}{}_2$, $\vec{m}{}_3$ are 
3-component column vectors. All of them depend on the spacetime
coordinates, via the given background fields. In addition, the 
$\boldsymbol{L}^{\rho}$ depend on $\vec{v}$ and the $\vec{m}{}_{\rho}$ depend on 
all dynamical variables $(\vec{E},\vec{B},\vec{v})$, but not on their 
derivatives. Hence, (\ref{eq:evovec}) is a quasi-linear system of 
first-order differential equations for $(\vec{E},\vec{B},\vec{v})$.

We can choose our coordinates such that, at the fixed spacetime point $P$,
we have $(g_{\mu \nu}) = \mathrm{diag}(-1,1,1,1)$ and 
$\Gamma _{\mu \nu}^{\sigma} = 0$. Then we can read from (\ref{eq:evo4}),
(\ref{eq:evo5}) and (\ref{eq:evo6}) that, at this point, 
\begin{gather}\label{eq:L}
\boldsymbol{L}^{\rho} \, = \, 
\begin{pmatrix}
-\vec{v} \otimes \vec{e}{\,}^{\rho} & - \boldsymbol{A}^{\rho} & \boldsymbol{0} \\
\boldsymbol{A}^{\rho} & \boldsymbol{0} & \boldsymbol{0} \\
\boldsymbol{0} & \boldsymbol{0} & -  v^{\rho} \, \boldsymbol{1} \\
\end{pmatrix}
\end{gather}
where 
%$\boldsymbol{1}$ is the $3 \times 3$ unit matrix, $\boldsymbol{0}$ is the
%$3 \times 3$ zero matrix,
\begin{gather}\label{eq:e}
\vec{e}{\,}^1 \, = \, \begin{pmatrix} 1 \\ 0 \\ 0 \end{pmatrix} \; , \qquad
\vec{e}{\,}^2 \, = \, \begin{pmatrix} 0 \\ 1 \\ 0 \end{pmatrix} \; , \qquad
\vec{e}{\,}^3 \, = \, \begin{pmatrix} 0 \\ 0 \\ 1 \end{pmatrix} \; , 
\\[0.2cm]
\label{eq:A}
\boldsymbol{A}^1 \, = \, 
\begin{pmatrix} 0 & 0 & 0 \\ 0 & 0 & 1 \\ 0 & -1 & 0 \end{pmatrix} \; , \qquad
\boldsymbol{A}^2 \, = \, 
\begin{pmatrix} 0 & 0 & -1 \\ 0 & 0 & 0 \\ 1 & 0 & 0 \end{pmatrix} \; , \qquad
\boldsymbol{A}^3 \, = \, 
\begin{pmatrix} 0 & 1 & 0 \\ -1 & 0 & 0 \\ 0 & 0 & 0 \end{pmatrix} \; .
\end{gather}
The $\vec{m}{}_{\rho}$ are given, at the point $P$, by
\begin{gather}\label{eq:m}
\vec{m}{}_1 \, = \, \vec{J}{}^e \, + \, J_0^e \vec{v}  \; , \qquad
\vec{m}{}_2 \, = \, \vec{0}  \; , \qquad
\vec{m}{}_3 \, = \, \dfrac{q}{m} \, 
\sqrt{ 1 -  v^{\tau}v^{\lambda} \delta _{\tau \lambda} } \,
\big( \vec{E} + \varepsilon_{\mu \nu \sigma} v^{\mu}B^{\nu} 
\vec{e}{\,}^{\sigma} -  v^{\rho} E_{\rho} \vec{v} \big)
\; .
\end{gather}
If $(M,g)$ is Minkowski spacetime, and if we use standard 
Minkowski coordinates, (\ref{eq:L}) and (\ref{eq:m}) hold not only at one point 
but everywhere.

%---------------------------------------------------------------------------------------------------
\section{Strong hyperbolicity of the Maxwell-Lorentz equations}\label{sec:strhyp}

In this section we will demonstrate that the initial-value problem for the
evolution equations of the Maxwell-Lorentz system is well-posed. If we assume
that the background fields are analytic, and that we give analytic initial 
data for $(\vec{E}, \vec{B}, \vec{v} )$, local existence and uniqueness of
a solution follows from the Cauchy-Kowalevsky theorem; this was the result
found by Parrott \cite{Parrott1987}. (Parrott only considered the case that
$(M,g)$ is Minkowski spacetime and that $J^e$ vanishes, but the result 
carries over immediately to our more general case.) For the reasons outlined
in the introduction, we want to have an existence and uniqueness statement
for non-analytic data. To that end we have to recall some terminology and
some results from the theory of partial differential equations.

For a quasi-linear system of first-order partial differential equations of the
form (\ref{eq:evovec}), the left-hand side of (\ref{eq:evovec}) is called the
\emph{principal part}. Existence and uniqueness of solutions is completely 
determined by the principal part and hence by the matrices $\boldsymbol{L}^{\rho}$.
(\ref{eq:evovec}) is called \emph{hyperbolic} if for any $\vec{p} = p_{\rho}
\vec{e}{\,}^{\rho} \in \mathbb{R}^3$ the eigenvalues of the $9 \times 9$ matrix 
$p_{\rho} \boldsymbol{L}^{\rho}$ are real. (\ref{eq:evovec}) is called 
\emph{strongly hyperbolic} or \emph{symmetrisable} if, in addition, for each 
eigenvalue the geometric multiplicity equals the algebraic multiplicity, i.e., if 
$p_{\rho} \boldsymbol{L}^{\rho}$ has nine linearly independent real eigenvectors.
An equivalent condition is that there exists a matrix $\boldsymbol{H}(\vec{p})$ 
such that $\boldsymbol{H}(\vec{p}) p_{\rho} \boldsymbol{L}^{\rho}$ is symmetric. 

According to a fundamental theorem from the theory of partial differential equations 
(see, e.g. Taylor \cite{Taylor1991}, Theorem 5.2.D), strong hyperbolicity guarantees 
that the initial-value problem is well-posed for Sobolev $H^s$ data. (A function 
is of class $H^s$ if its $s^{\mathrm{th}}$ derivative exists almost everywhere 
and is locally square-integrable.) In general, $s$ must be chosen greater than 
$(n+2)/2$, where $n$ is the dimension of the initial hypersurface. If adapted
to our case, the theorem can be stated in the following way.

Assume that (\ref{eq:evovec}) is strongly hyperbolic, with the matrices
$\boldsymbol{L}^{\rho}$ and the vectors $\vec{m}_{\rho}$ smoothly depending
on the spacetime coordinates and on the dynamical variables $(\vec{E},
\vec{B}, \vec{v})$. Choose initial data of class $H^s$ for 
$(\vec{E}, \vec{B} ,\vec{v})$ on the hypersurface $x^0=$constant 
through the spacetime point $P$, where $s \ge 3$. Then 
\begin{itemize}
\item
a solution to (\ref{eq:evovec}) with the prescribed initial data exists
on a neighborhood of $P \,$;
\item
on this neighborhood, the solution is uniquely determined by the initial 
data;
\item
the solution depends continuously, with respect to the $H^s$ topology,
on the initial data.
\end{itemize}

It is now our goal to demonstrate that the evolution equations of the 
Maxwell-Lorentz system are strongly hyperbolic. If this result has been
established, we know that the initial-value problem is well-posed,
provided that our background fields $g_{\mu \nu}$ and $J_{\mu}^e$ 
depend smoothly on the spacetime coordinates and the initial data are
of class $H^s$ with $s \ge 3$. 

As strong hyperbolicity is an algebraic property that is to be checked
pointwise, we may assume that the matrices $\boldsymbol{L}^{\rho}$ are 
given by (\ref{eq:L}), hence
\begin{gather}\label{eq:pL}
p_{\rho} \boldsymbol{L}^{\rho} \, = \, 
\begin{pmatrix}
-\vec{v} \otimes \vec{p} & - p_{\rho} \boldsymbol{A}^{\rho} & \boldsymbol{0} \\
p_{\rho} \boldsymbol{A}^{\rho} & \boldsymbol{0} & \boldsymbol{0} \\
\boldsymbol{0} & \boldsymbol{0} & -  p_{\rho} v^{\rho} \, \boldsymbol{1} \\
\end{pmatrix}
\; .
\end{gather}
We have to show that this matrix has nine linearly independent real 
eigenvectors for each $\vec{p}=p_{\rho}\vec{e}{\,}^{\rho} \in \mathbb{R}^3$. 
We first determine the eigenvalues. $p_0$ is an eigenvalue of 
$p_{\rho}\boldsymbol{L}^{\rho}$ if the characteristic equation
\begin{equation}\label{eq:charac}
\mathrm{det} \big( p_0 \boldsymbol{1} - p_{\rho} \boldsymbol{L}^{\rho} \big)
\, = \, 0
\end{equation}
is satisfied. If this equation holds, $p_idx^i=p_0dx^0+p_{\mu}dx^{\mu}$ is 
called a \emph{characteristic covector} of the evolution equations. Upon
inserting $p_{\rho} \boldsymbol{L}^{\rho}$ from (\ref{eq:pL}) into
(\ref{eq:charac}), the determinant can be calculated with the help of 
some elementary algebra. As a result of this calculation, the characteristic 
equation reads
\begin{equation}\label{eq:p0}
(p_0^2-\delta^{\mu \nu}p_{\mu}p_{\nu})^2 
\, (p_0 + p_{\rho}v^{\rho})^4 \, p_0
\, = \, 0 \; , 
\end{equation}
so the eigenvalues are $\sqrt{\delta^{\mu \nu}p_{\mu} p _{\nu}}$,
$-\sqrt{\delta^{\mu \nu} p_{\mu} p _{\nu}}$, $-p_{\rho}v^{\rho}$ and
0. As all eigenvalues are real, this demonstrates that the evolution equations 
are hyperbolic. 

We will now prove that they are even strongly hyperbolic. To that end we
have to determine the corresponding eigenvectors. As the case $\vec{p}
= \vec{0}$ is trivial, we may assume that $\vec{p} \neq \vec{0}$. 
Owing to the homogeneity of the eigenvalue equation it suffices to 
consider the case that $\delta ^{\mu \nu} p_{\mu} p_{\nu} = 1$. It is
then straight-forward to verify  that
\begin{equation}\label{eq:eigen1}
\mathcal{E} _1 \, = \, \left\{ \left.
\begin{pmatrix}
\vec{u} \\ - p_{\rho}  \boldsymbol{A} ^{\rho} \vec{u} \\ \vec{0}
\end{pmatrix}
\,  \right| \, 
\vec{u} = u_{\rho} \vec{e}{\,}^{\rho} \in \mathbb{R}^3 \, , \quad
\delta ^{\mu \nu} u_{\mu} p_{\nu} = 0  
\; \right\}
\end{equation}
is a two-dimensional eigenspace of the eigenvalue $p_0 = 1$,
\begin{equation}\label{eq:eigen2}
\mathcal{E} _2 \, = \, \left\{ \left.
\begin{pmatrix}
- p_{\rho} \boldsymbol{A} ^{\rho} \vec{w} \\ \vec{w} \\ \vec{0}
\end{pmatrix}
\, \right| \, 
\vec{w} = w_{\rho} \vec{e}{\,}^{\rho} \in \mathbb{R}^3 \, , \quad
\delta ^{\mu \nu} w_{\mu} p_{\nu} = 0  
\; \right\}
\end{equation}
is a two-dimensional eigenspace of the eigenvalue $p_0 = - 1$, 
\begin{equation}\label{eq:eigen3}
\mathcal{E} _3 \, = \, \left\{ \left.
\begin{pmatrix}
\alpha \, \big( \, \vec{p} - p_{\rho}v^{\rho} \vec{v} \, \big) \, \\ 
-\alpha \, p_{\rho} \boldsymbol{A} ^{\rho} \vec{v} \,
\\ \vec{z}
\end{pmatrix}
\, \right| \, 
\alpha \in \mathbb{R} \, , \quad \vec{z} \in \mathbb{R}^3 \; \right\}
\end{equation}
is a four-dimensional eigenspace of the eigenvalue $p_0=-p_{\rho}v^{\rho}$,
and
\begin{equation}\label{eq:eigen4}
\mathcal{E} _4 \, = \, \left\{ \left.
\begin{pmatrix}
\vec{0} \\ \beta \vec{p} \\ \vec{0}
\end{pmatrix}
\, \right| \,
\beta \in \mathbb{R} \right\}
\end{equation}
is a one-dimensional eigenspace of the eigenvalue $p_0=0$. The eigenspaces 
$\mathcal{E} _1$, $\mathcal{E} _2$, $\mathcal{E} _3$ and $\mathcal{E} _4$
span the whole space, even in the case of further degeneracy (i.e., even 
if $p_{\rho}v^{\rho}=0$ or $p_{\rho} v^{\rho} = \pm 1$). This demonstrates 
that the evolution equations are strongly hyperbolic.

This concludes the local existence-and-uniqueness proof for 
solutions $(\vec{E},\vec{B},\vec{v})$ to the evolution equations.
For solving the full set of Maxwell-Lorentz equations, we have
to single out those solutions to the evolution equations that
satisfy the constraints. As a consequence, not all eigenvectors
in  $\mathcal{E} _1$, $\mathcal{E} _2$, $\mathcal{E} _3$ and
$\mathcal{E} _4$ are associated with solutions of the full set
of Maxwell-Lorentz equations. The condition $n>0$ does not 
restrict the eigenvectors. However, equation  (\ref{eq:con3}) 
reduces the eigenspaces according to  
\begin{equation}\label{eq:eigenred}
\hat{\mathcal{E}}{} _A \, = \, \left\{ \left.
\begin{pmatrix}
\vec{E}{} _0 \\ \vec{B}{}_0 \\ \vec{v}{}_0
\end{pmatrix}
\in \mathcal{E} _A
\, \right| \,
p_{\mu}B_0^{\mu}= 0 \, \right\}
\: , \quad A=1,2,3,4 \, .
\end{equation}
By inspection we find that $\hat{\mathcal{E}}{}_1 = \mathcal{E}_1$,
$\hat{\mathcal{E}}{}_2 = \mathcal{E}_2$, $\hat{\mathcal{E}}{}_3 
= \mathcal{E}_3$ whereas $\hat{\mathcal{E}}{}_4$ consists of 
the zero-vector only, so $p_0=0$ is no longer an eigenvalue if the
constraints are taken into account. This reduces the characteristic
equation of the evolution equations (\ref{eq:p0}) to the characteristic
equation of the full set of Maxwell-Lorentz equations,
\begin{equation}\label{eq:p0red}
(p_0^2-\delta^{\mu \nu}p_{\mu}p_{\nu})^2 
\, (p_0 + v^{\rho}p_{\rho})^4 \, = \, 0 \; . 
\end{equation}
In physical terms, (\ref{eq:p0red}) gives us the three branches
of the \emph{dispersion relation},
\begin{equation}\label{eq:dispersion}
p_0 = \sqrt{\delta^{\mu \nu}p_{\mu}p_{\nu}} \, , \quad
p_0 = -\sqrt{\delta^{\mu \nu}p_{\mu}p_{\nu}} \, , \quad
p_0 = - v^{\rho}p_{\rho} \; .
\end{equation}
From the corresponding eigenspaces we read that the first
two branches are associated with the (past-oriented and
future-oriented, respectively) propagation of the electromagnetic field, 
whereas the third branch is associated with the propagation of the 
dust. If we calculate the group velocity (or ray velocity)
\begin{equation}\label{eq:group}
\mathrm{v}^{\mu} _{\mathrm{gr}} \, = \, 
\dfrac{\partial (-p_0)}{\partial p_{\mu}}  \, ,
\end{equation}
for each branch of the dispersion relation we see that the electromagnetic 
field can propagate in any spatial direction with the vacuum speed of light 
( $\mathrm{v}^{\mu} 
_{\mathrm{gr}} \mathrm{v}^{\nu} _{\mathrm{gr}} 
\delta _{\mu \nu} = 1$), whereas the dust 
propagates with $\mathrm{v}^{\mu} _{\mathrm{gr}} = v^{\mu}$. 

%---------------------------------------------------------------------------------------------------
\section{Example: Spherically symmetric solutions on Minkowski spacetime}\label{sec:example}
Knowing that the initial-value problem admits a unique solution is one thing, actually
determining this solution is another. In most cases one can find the solution only 
numerically. However, in some cases of high symmetry it is actually possible to work out 
some features of the solution explicitly. Parrot \cite{Parrott1987} considers two such cases: (i) 
spherically symmetric solutions on Minkowski spacetime, and (ii) solutions on Minkowski 
spacetime that depend only on one of the three spatial Cartesian coordinates $(x,y,z)$. 
The first case is concerned with expanding or collapsing shells of charge, and the second 
case with moving ``walls'' of charge. Burton, Gratus and Tucker also considered the first case 
\cite{BurtonGratusTucker2007a} and the second case \cite{BurtonGratusTucker2007}.

Here we want to revisit spherically symmetric solutions. Using Parrott's treatment
of this case as a general guide, we will get some results that go beyond his findings.
In particular, we will solve the differential equations for the flow lines of the dust
explicitly, and we will get some additional insight into the qualitative features of the 
solutions. Although spherically
symmetric fields are somewhat over-idealised, we find it instructive to study this
case. As we will see, it illustrates the way in which the constraints have to be 
built in and it demonstrates how caustics are formed, thereby preventing the solution
from being global.

We assume that the background metric is the Minkowski metric and that the external
current vanishes, $J_i^e=0$. In spherical polar coordinates $(x^0,x^1,x^2,x^3)=(t,r,
\vartheta , \varphi)$, the metric reads
\begin{equation}\label{eq:polar}
g_{ij}dx^idx^j \, = \, - \, dt^2 \, + \, dr^2 \, + \, r^2 \, \big( \, \mathrm{sin} ^2
\vartheta \, d \varphi ^2 \, + \, d \vartheta ^2 \, \big) \; .
\end{equation}
For spherically symmetric situations, the constraints equation (\ref{eq:con3}) can 
hold with a non-zero magnetic field only if there is a magnetic monopole at $r=0$. 
We exclude the existence of magnetic monopoles and, thus, assume that $B^{\sigma}=0$.
We are then left with only two dynamical variables,
\begin{gather}
\label{eq:Epolar}
E_{\mu} dx^{\mu} \, = \, E(r,t) dr \; ,
\\
\label{eq:vpolar}
v^{\mu} \partial _{\mu} \, = \, v(r,t) \, \partial _r \; ,
\end{gather}
and the evolution equations (\ref{eq:evo4}), (\ref{eq:evo5}) and 
(\ref{eq:evo6}) are identically satisfied, apart from the following two
scalar equations,
\begin{gather}
\label{eq:evopolar1}
\dfrac{\partial E(r,t)}{\partial t} \, + \, 
v(r,t) \, \dfrac{\partial E(r,t)}{\partial r}
\, = \, - \, \dfrac{2}{r} \, v(r,t) \, E(r,t) \; ,
\\[0.2cm]
\label{eq:evopolar2}
\dfrac{\partial v(r,t)}{\partial t} \, + \, 
v(r,t) \, \dfrac{\partial v(r,t)}{\partial r}
\, = \, \dfrac{q}{m} \, \sqrt{1-v(r,t)^2 \,}^3 \, E(r,t)  \; ,
\end{gather}
The constraints (\ref{eq:con3}) and (\ref{eq:con4}) reduce to a single
scalar equation,
\begin{gather}
\label{eq:conpolar}
q \, n(r,t) \, = \, \sqrt{1-v(r,t)^2} \, 
\Big( \, \dfrac{\partial E(r,t)}{\partial r}
\, + \dfrac{2}{r} \, E(r,t) \, \Big) \; ,
\end{gather}
which gives the charge density $q \, n(r,t)$ in terms of the dynamical variables
$E(r,t)$ and $v(r,t)$. Note that the electric field is to be interpreted as the 
field produced by the charged dust plus, possibly, a source-free field. The only 
spherically symmetric electric field that is source-free, on a spherical shell
$r_1 < r < r_2$, is a Coulomb field; i.e., it can be realised by a point
charge at $r=0$.

To solve the system of equations (\ref{eq:evopolar1}), (\ref{eq:evopolar2}),
and (\ref{eq:conpolar}) we would have to give
initial values $E(r,t_0)$ and $v(r,t_0)$ on a certain interval 
$r_1<r<r_2$. With such initial data given, our existence and uniqueness 
result guarantees that there is a unique local solution to (\ref{eq:evopolar1}) 
and (\ref{eq:evopolar2}). From all possible initial values we would then have 
to discard all those for which (\ref{eq:conpolar}) gives a negative number 
density $n(r,t)$ on the initial hypersurface.

The latter constraint is a bit awkward. It would be more convenient
if we could give the initial data in terms of $n(r,t_0)$ directly. Also,
if we want to explicitly integrate (\ref{eq:evopolar1}) and (\ref{eq:evopolar2})
we have to face the problem that these two equations are coupled. It would 
be nicer to have an equation for either the electric field or the flowlines of 
the dust alone. Both goals can, indeed, be achieved if we rearrange our 
system of equations appropriately. To that end we first define
\begin{equation}\label{eq:Q}
Q(r,t) \, = \, 4 \, \pi \, r^2 \, E(r,t) \; .
\end{equation}
This definition follows Parrott, adapted to our choice of units. (Parrott has a
factor $4 \pi$ in Maxwell's equations, we don't.) Then (\ref{eq:conpolar}) can 
be rewritten as
\begin{equation}\label{eq:nQ}
q \, n(r,t) \, = \, \dfrac{\sqrt{1-v(r,t)^2 \,}}{4 \, \pi \, r^2} \,
\dfrac{\partial Q(r,t)}{\partial r} \; .
\end{equation}
Upon integration, this yields 
\begin{equation}\label{eq:Qint}
Q(r,t) \, = \, Q(r_0,t) \, + \, 4 \, \pi \, q \, \int _{r_0} ^r
\, \dfrac{\, n(s,t) \, s^2 \, ds \,}{\sqrt{1-v(s,t)^2 \,}} 
\end{equation}
where $r_0$ is any chosen radius value. As $q \, n(r,t)$ is the proper charge density,
$q \, n(r,t)/\sqrt{1-v(r,t)^2}$ is the charge density in the chosen inertial frame. Hence we 
read from equation (\ref{eq:Qint}) that $Q(r,t) - Q(r_0,t)$ is the total charge in 
the shell between radius $r_0$ and radius $r$ at time $t$. After replacing the 
original pair of dynamical variables $(E,v)$ by $(Q,v)$, the evolution 
equations (\ref{eq:evopolar1}) and (\ref{eq:evopolar2}) read
\begin{gather}
\label{eq:evopolar3}
\dfrac{\partial Q(r,t)}{\partial t} \, + \, 
v(r,t) \, \dfrac{\partial Q(r,t)}{\partial r} \, = \, 0 \; ,
\\[0.2cm]
\label{eq:evopolar4}
\dfrac{\partial v(r,t)}{\partial t} \, + \, 
v(r,t) \, \dfrac{\partial v(r,t)}{\partial r} \, = \, \dfrac{q}{m}
\, \dfrac{\sqrt{1-v(r,t)^2 \,}^{\, 3}\,}{4 \, \pi \, r^2} \, Q(r,t) \; .
\end{gather} 
If we evaluate these equations along a flow line $\big(r(t), \varphi (t) = \varphi _0
, \vartheta (t) = \vartheta _0 \big)$ of the dust, such that $v \big( r(t),t \big) = d r(t)
/dt$, we find
\begin{gather}
\label{eq:evopolar5}
\dfrac{d Q \big( r(t),t \big)}{d t}  \, = \, 0 \; ,
\\[0.2cm]
\label{eq:evopolar6}
\dfrac{d v \big( r(t),t \big)}{d t}  \, = \, \dfrac{q}{m} \, 
\dfrac{\sqrt{1-v \big( r (t),t \big) ^2 \,}^{\, 3} \,}{4 \, \pi \, r^2} 
\, Q \big( r(t) ,t \big)
 \; .
\end{gather} 
The first equation says that $Q$ is conserved along each flow line of the
dust. With this information at hand, the second equation can be viewed
as a second-order differential equation for the flow lines alone. We can, thus,
solve our system of equations in the following way.

Fix a time $t_0$ and a radius value $r_0$. Prescribe, on an interval
$r_1<r<r_2$, initial values
\begin{gather}
\label{eq:inin}
n(r,t_0) \, = \, n_0(r) \; , \qquad n_0(r) \, > \, 0 \; ,
\\[0.2cm]
\label{eq:iniv}
v(r,t_0) \, = \, v_0(r) \; , \qquad v_0(r)^2 \, < \, 1 \; ,
\\[0.2cm]
\label{eq:iniE0}
E(r_0,t_0) \, = \, E_0 \; .
\end{gather}
The freedom of choosing a value for $E_0$ corresponds to the freedom of adding
a Coulomb field, i.e., a point charge at $r=0$. 

The data (\ref{eq:inin}), (\ref{eq:iniv}) and (\ref{eq:iniE0}) determine initial 
values for the function $Q$, namely
\begin{equation}\label{eq:Qini}
Q(r,t_0) \, = \, 4 \, \pi \, r_0^2 \, E_0 \, + \, 4 \, \pi \, q \, 
\int _{r_0} ^r  \, \dfrac{\, n_0(s) \, s^2 \, ds \,}{\sqrt{1-v_0(s)^2 \,}} \; .
\end{equation}
Then (\ref{eq:evopolar6}) gives us a second-order differential equation for
the flow lines of the dust,
\begin{equation}\label{eq:diffr}
\dfrac{d^2r(t)}{dt^2} \, = \, \dfrac{q}{m} \, \Big\{ \, 1 \, - \,
\Big( \dfrac{dr(t)}{dt} \Big) ^2 \, \Big\} ^{3/2} \, 
\dfrac{Q(r',t_0)}{r(t)^2} \; ,
\end{equation}
that has to be solved with initial conditions 
\begin{equation}\label{eq:inir}
r(t_0)=r' \qquad \mathrm{and} \qquad \dfrac{dr}{dt} \big(t_0 \big) \, = \,
v_0 (r')
\end{equation}
for all $r_1<r'<r_2$. If we have determined the flow lines, our dynamical problem
is essentially solved. It is true that knowing the flow lines does not allow us 
to write closed form expressions for the other dynamical variables.
However, all the relevant information is then known: The function $Q$, which
determines the charge in any spherical shell, is fixed by being constant along 
the flow lines; thereupon the electric field is fixed by (\ref{eq:Q}). The solution 
exists and is unique until the flow lines begin crossing over and forming caustics.

So our problem has been reduced, in essence, to solving the differential
equation (\ref{eq:diffr}). This can actually be done explicitly.  
After multiplying both sides with $dr(t)/dt$, (\ref{eq:diffr}) can be 
rewritten as
\begin{equation}\label{eq:firstint}
\dfrac{d}{dt} \Big( \, \Big\{ 1 - \Big( \dfrac{dr(t)}{dt} \Big)^2 \Big\}
^{-1/2} \, \Big) \, = \, - \, \dfrac{\, q \, Q(r',t_0)\,}{m} \, \dfrac{d}{dt}
\Big( \, \dfrac{1}{r(t)} \, \Big) \; .
\end{equation}
This gives us a first integral,
\begin{equation}\label{eq:intA}
\Big\{ 1 - \Big( \dfrac{dr(t)}{dt} \Big)^2 \Big\}
^{-1/2} \, + \, \dfrac{\, q \, Q(r',t_0) \,}{m \, r(t)} \, = \, A(r') \; .
\end{equation}
The value of the constant of motion $A(r')$ is determined by evaluating
(\ref{eq:intA}) at $t=t_0$,
\begin{equation}\label{eq:valA}
\big\{ 1 - v_0(r')^2 \big\}^{-1/2} \, + \, \dfrac{\, q \, Q(r',t_0) \,}{m \, r'} \, = \, A(r') \; .
\end{equation}
Equation (\ref{eq:intA}) has the interesting consequence that a dust particle
can reach the origin, $r(t)=0$, only if it arrives there at the speed of light, 
$\big( dr(t) /dt \big) ^2 = 1$.
The only exception are trajectories with $Q(r',t_0) =0$ which, by (\ref{eq:diffr}), are
straight lines. Note that trajectories with $qQ(r',t_0) > 0$ are accelerating outwards
and trajectories with $qQ(r',t_0) < 0$ are accelerating inwards. 

The first-order differential equation (\ref{eq:intA}) can be solved by separation of 
variables. We find
\begin{gather}\label{eq:pos}
\pm \, t \; = \; 
\dfrac{\, q \, Q(r',t_0) \,}{m Z(r')^{3/2}} \,
\mathrm{ln} \left(
2 \, \sqrt{Z(r')} \, 
\sqrt{ Z(r')  \, r^2 \, - \, 2 \, 
\dfrac{\, q \, Q(r',t_0) \,}{m} \, A(r') \, r \, + \,
\frac{\, q^2 \, Q(r',t_0)^2 \, }{m^2}} \right.
\\[0.2cm]
\nonumber
\left.
 + \, 2  Z(r') \, r  \, - \, 
2 \, \frac{\, q \, Q(r',t_0) \,}{m}A(r') \right) \; + \; B(r')
\end{gather}
if $Z(r')= A(r')^2-1 >0$, 
\begin{equation}\label{eq:neg}
\pm \, t \; = \; 
\dfrac{\, q \, Q(r',t_0) \, }{m \, \big( -Z(r') \big)^{3/2}} \; 
\mathrm{arcsin} \,
\left(
\dfrac{2 \, Z(r') \, r \, - \, 2 \, 
\dfrac{
\, q \, Q(r',t_0) \, }{m} \, A(r')
}{
\sqrt{\, 4 \, \dfrac{A(r')^2 \, q^2 \, Q(r',t_0)^2}{m^2} \, - \, 
4 \, Z(r')  \, \dfrac{\, q^2 \, Q(r',t_0)^2}{m^2}}
}
\right) \; + \; B(r')
\end{equation}
if $Z(r')=A(r')^2-1<0$, and
\begin{equation}\label{eq:null}
\pm \, t \; = \; 
\sqrt{\, - \, 2 \, 
\dfrac{\, q \, Q(r',t_0) \,}{m} \, A(r') \, r \, + \,
\dfrac{\, q^2 \, Q(r',t_0)^2}{m^2}} \, 
\Big( \dfrac{2}{\, 3\, A(r') \,} \, - \, 
\dfrac{\, m \, r \, }{\, 3 \, q \, Q(r',t_0) \,} \Big) 
\; + \; B(r') 
\end{equation}
if $Z(r')=A(r')^2-1=0$. In all three cases, the integration constant $B(r')$ is
determined by setting $t=t_0$ and $r=r'$, and the sign on the 
left-hand side must be chosen positive for outward going and 
negative for inward going parts of the trajectory.

For any specific choice of the initial conditions, (\ref{eq:pos}), (\ref{eq:neg})
and (\ref{eq:null}) give us the flow lines of the dust which can
then be plotted in an $(r,t)$ spacetime diagram. For the sake
of illustration, we choose the following initial conditions on 
the interval $0 < r < r_0$,
\begin{gather}
\label{eq:n0}
n (r, t_0) \, = \,
a \, (r_0-r)^2 \; ,
\\[0.2cm]
\label{eq:v0}
v(r,t_0) \, = \, 0 \; ,
\\[0.2cm]
\label{eq:E0}
E(r_0,t_0) \, = \, E_0 \; ,
\end{gather}
where $a$ is a positive constant and $E_0$ is a constant that
may be chosen positive, negative, or zero. In this case our system 
describes a charged ball of radius $r_0$ that is initially at rest; in 
addition, there is a point charge at the origin whose value depends 
on the chosen value of $E_0$.

\begin{figure}[h]
    \psfrag{r0}{$\,$ \hspace{-0.25cm}  $r _0$} 
    \psfrag{r}{$\,$ \hspace{-0.0cm} $r$} 
    \psfrag{t}{$\,$ \hspace{-0.9cm} $t$} 
    \psfrag{O}{$\,$ \hspace{0cm} $$} 
\centerline{\epsfig{figure=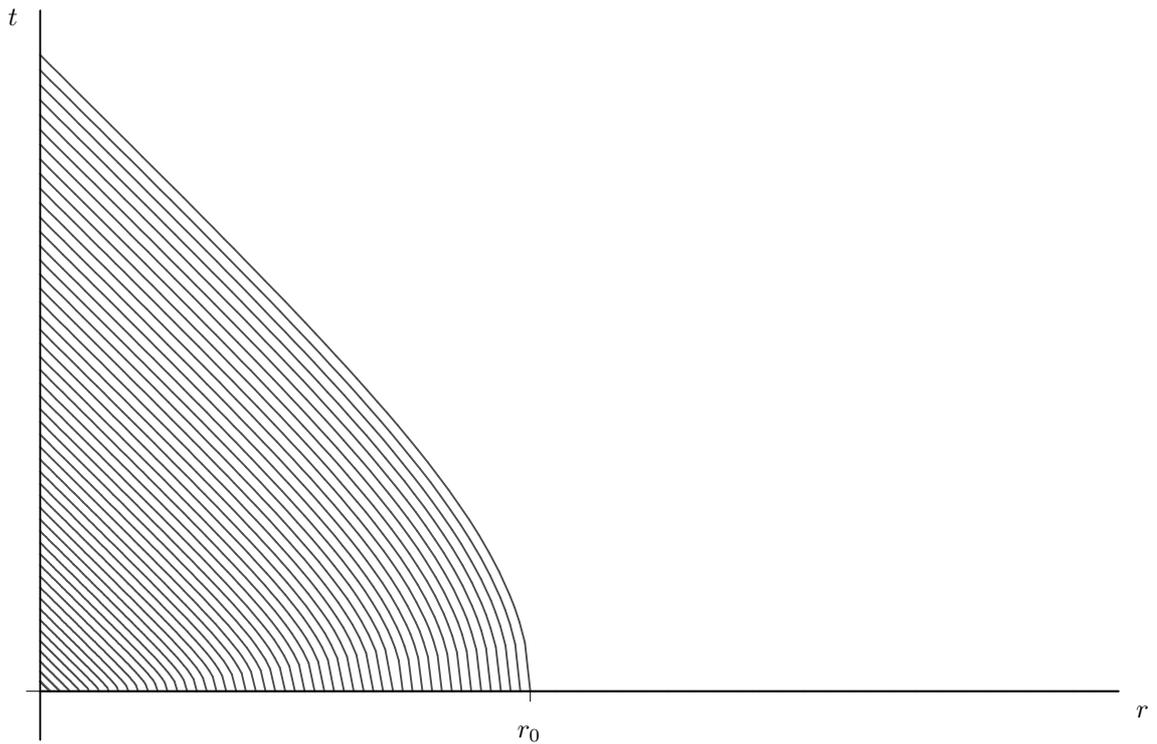, width=15cm}}
\caption{If $qE_0$ is negative, $qQ(r,t_0)$ is negative on 
the whole interval $0<r<r_0$. In this case there is a point 
charge at the centre whose charge has the opposite sign 
as the charge of the dust. The Coulomb attraction of this 
point charge is so strong that it overcomes the Coulomb 
repulsion of the dust particles and sucks the whole ball of 
charge into the origin in  a finite time. All dust particles 
arrive at the origin with the speed of light. Although the 
initial condition for the dust's density, $n(r,t_0)$, admits 
a regular extension into the origin, the density is singular 
at the origin for all later times. If one assumes that
the dust particles can pass through each other when
they meet at the origin, the ball would re-expand
after its collapse.
\protect\label{fig:Figure1}}
\end{figure}

\begin{figure}[h]
    \psfrag{r0}{$\,$ \hspace{-0.25cm}  $r _0$} 
    \psfrag{r}{$\,$ \hspace{-0.0cm} $r$} 
    \psfrag{t}{$\,$ \hspace{-0.9cm} $t$} 
    \psfrag{O}{$\,$ \hspace{0cm} $$} 
\centerline{\epsfig{figure=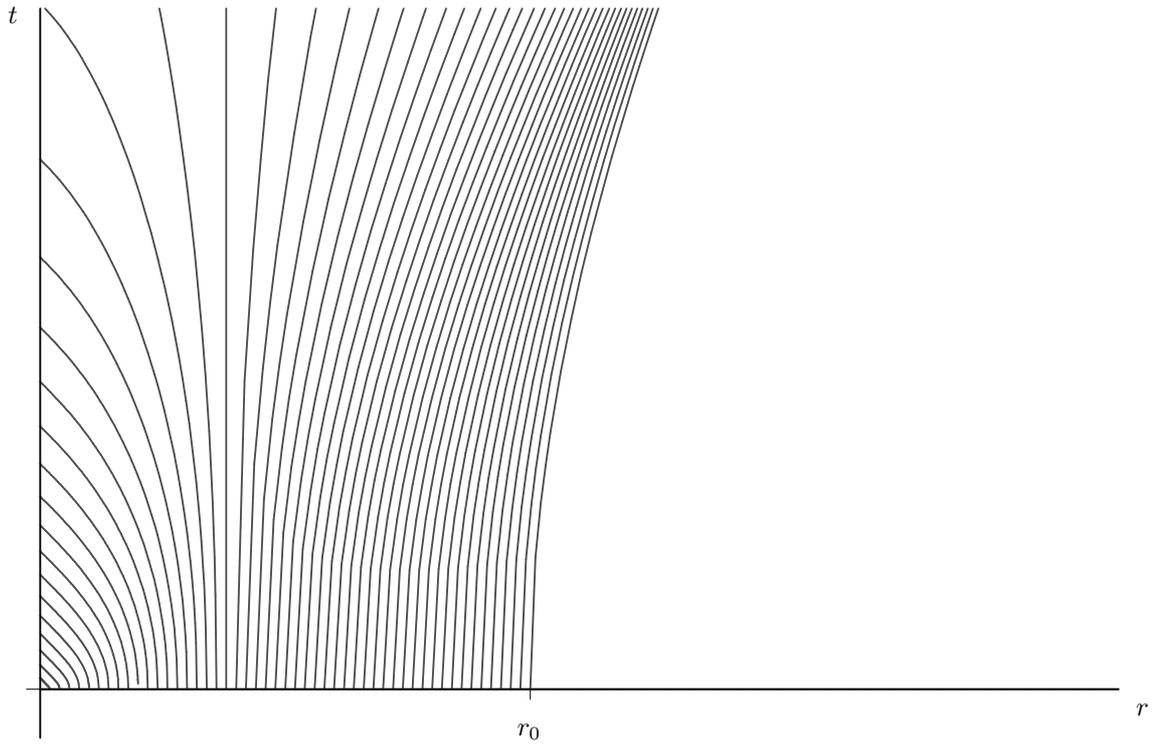, width=15cm}}
\caption{If $0<qE_0< q^2ar_0^3/30$, there is a radius 
value $r$ between 0 and $r_0$ for which $Q(r,t_0)$ becomes 
zero. At this radius, the ball is stationary. The outer part 
expands because here the Coulomb repulsion of the dust 
particles is stronger than the Coulomb attraction of the point 
charge towards the origin; the inner part collapses because 
here it is vice versa. Again, any dust particle that reaches 
the origin arrives there at the speed of light, and the 
density $n(r,t)$ is singular at $r=0$ for all $t>t_0$.
\protect\label{fig:Figure2}}
\end{figure}

\begin{figure}[h]
    \psfrag{r0}{$\,$ \hspace{-0.25cm}  $r _0$} 
    \psfrag{r}{$\,$ \hspace{-0.0cm} $r$} 
    \psfrag{t}{$\,$ \hspace{-0.9cm} $t$} 
    \psfrag{O}{$\,$ \hspace{0cm} $$} 
\centerline{\epsfig{figure=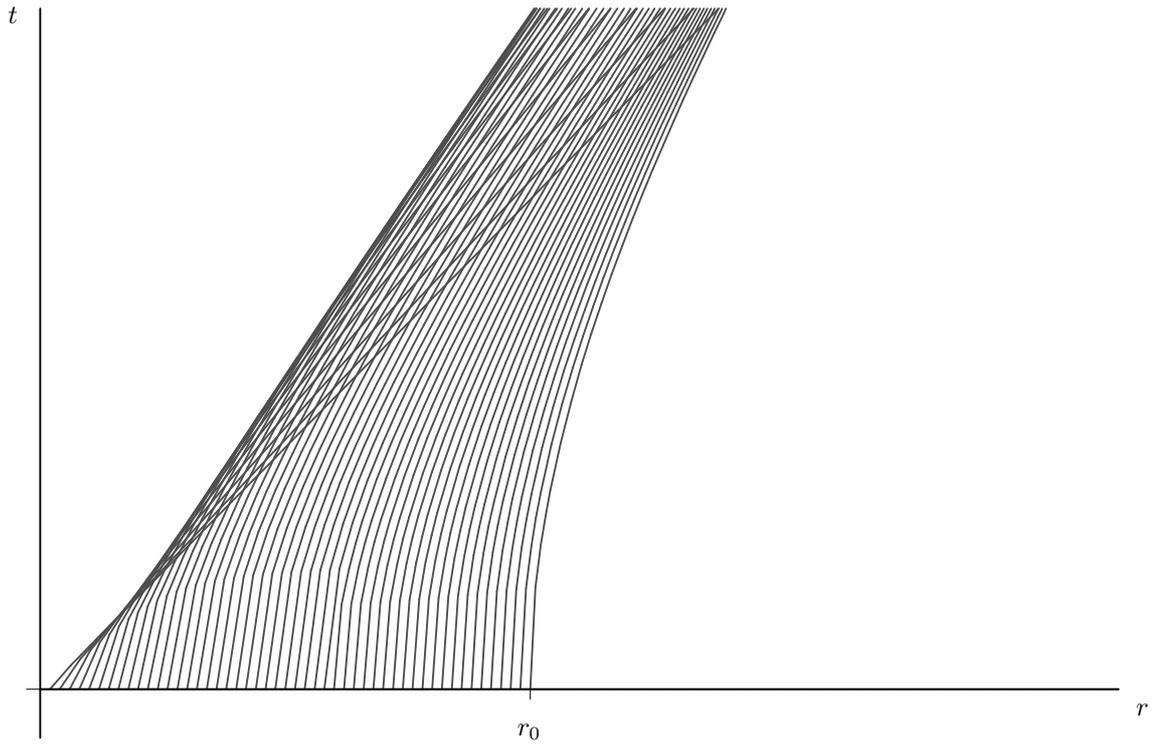, width=15cm}}
\caption{If $qE_0> q^2ar_0^3/30$, the point charge 
at the centre has the same sign as the charge of
the dust, so it enhances the tendency of the dust 
to expand outwards. The inner parts of the 
ball move faster than the outer parts and overtake 
them, thereby forming a caustic. The solution to the 
Maxwell-Lorentz equations breaks down when
flow lines cross over. As guaranteed by our general
theorem, the solution is well-defined and unique on
a neighborhood of the initial hypersurface. In this
case, the latter is a snapshot of a punctured ball.
\protect\label{fig:Figure3}}
\end{figure}

Figures \ref{fig:Figure1}, \ref{fig:Figure2} and \ref{fig:Figure3} show 
Maple plots of the flow lines in an $(r,t)$ spacetime diagram for 
three qualitatively different cases, depending on the chosen value 
of $E_0$.  As discussed in the captions of the figures, this example 
nicely demonstrates the role of the point charge at the origin (and, 
thereby, of the fact that in general the choice of initial values for 
the Maxwell-Lorentz equations includes the choice of a source-free 
electromagnetic field). It also illustrates how a caustic is formed 
(which is the main reason why, in general, existence and uniqueness 
of  a solution is guaranteed only locally near the initial hypersurface).

\newpage
$\,$

\newpage
$\,$

\newpage

$\,$

\newpage
%---------------------------------------------------------------------------------------------------
\section{Conclusion}\label{sec:con}
In this article we have demonstrated that
the initial-value problem of the Maxwell-Lorentz equations is
well-posed, and we have illustrated some qualitative
features of these equations by studying spherically symmetric 
solutions on Minkowski spacetime. In view of
applications, it is desirable to generalise our result in
two directions. Firstly, one would like to include a 
pressure term into the equations, thereby allowing
for situations where temperature and density are so
high that the model of a ``dust'' (or ``cold fluid'')
is no longer applicable. Secondly, one would like to 
consider mixtures of two or more fluids which are
dynamically coupled together. This would allow for
considering two-fluid models of plasmas, treating
both the electron component and the ion component
as dynamical fields. In particular the first generalisation
would require a considerable change in the mathematical
techniques. Whereas for our treatment of a dust it
was crucial that the density could be eliminated from
the evolution equations, this would no longer be true
for a fluid with pressure. One would have to use 
either the density or the pressure as an additional
dynamical variable, and to eliminate the other one 
with the help of an equation of state.   
One would also expect that a pressure term would 
change the behaviour of a collapsing ball of charge. 
In particular, the somewhat pathological behaviour 
near the centre, as discussed in Chapter 
\ref{sec:example},  is likely to be modified by a 
pressure term.

%---------------------------------------------------------------------------------------------------
\section*{Acknowledgment}
VP wishes to thank Nico Giulini for clarifying comments on 
the interpretation of solutions to the Maxwell-Lorentz equations.
VP has also profited from discussions on the subject with his 
Lancaster colleagues, in particular with David Burton, Jonathan 
Gratus and Robin Tucker.

\end{document}